\documentclass{scrartcl}

\usepackage[utf8]{inputenc}
\usepackage[T1]{fontenc}
\usepackage[margin=1.5cm]{geometry}
\usepackage[svgnames]{xcolor}
\usepackage{multicol,microtype,enumitem,lmodern}

\usepackage{mathtools,amssymb}

\usepackage[colorlinks=true,allcolors=DarkBlue]{hyperref}
\usepackage{cleveref}
\Crefname{equation}{Eq.}{Eqs.}

\usepackage{tikz}
\usetikzlibrary{patterns,shapes.misc}
\tikzset{cross/.style={cross out,draw=black,minimum size=2*#1},cross/.default=0.5}

\usepackage[backend=biber,sorting=none,citestyle=numeric-comp,giveninits=true,doi=false,date=year,url=false]{biblatex}
\renewbibmacro{in:}{}
\AtEveryBibitem{\clearfield{number}\clearfield{issue}}
\DeclareFieldFormat[article]{volume}{\mkbibbold{#1}}

\DeclareFieldFormat{pages}{\mkfirstpage{#1}}
\DeclareFieldFormat{eprint:arxiv}{\href{http://arxiv.org/abs/#1}{#1}}

\newcommand{\dif}{\mathrm{d}}
\newcommand{\normord}[1]{\vcentcolon\mathrel{#1}\vcentcolon}
\DeclareMathOperator{\Tr}{Tr}
\newenvironment{alignedeqn}{\begin{equation}\begin{aligned}}{\end{aligned}\end{equation}\ignorespacesafterend}

\title{Exact evolution equation for the effective potential}
\author{
    Christof Wetterich\\
    \small\href{mailto:c.wetterich@thphys.uni-heidelberg.de}{c.wetterich@thphys.uni-heidelberg.de}\\
    \small Institut für Theoretische Physik, Universität Heidelberg, Philosophenweg 16, 69120 Heidelberg, Germany}
\date{\small
	15 November 1992; revised 17 December 1992\\
    published in \href{http://www.sciencedirect.com/science/article/pii/037026939390726X}{Physics Letters B 301 (1993) 90-94}}

\begin{document}

\maketitle

\begin{abstract}
    We derive a new exact evolution equation for the scale dependence of an effective action. The corresponding equation for the effective potential permits a useful truncation. This allows one to deal with the infrared problems of theories with massless modes in less than four dimensions which are relevant for the high temperature phase transition in particle physics or the computation of critical exponents in statistical mechanics.
\end{abstract}

\begin{multicols}{2}

The average action $\Gamma_k$ \cite{wetterich1991average} is an effective action for averages of fields. The average is taken over a volume $\sim k^{-d}$ such that all degrees of freedom with momenta $q^2 > k^2$ are effectively integrated out. The average action is formulated in continuous space and is the analogue of the block spin action proposed earlier on a lattice \cite{wilson1971renormalization,wilson1974renormalization,kadanoff1966scaling}. Recently, a renormalization group improved one-loop equation has been derived \cite{wetterich1993average} for the average potential $U_k$. For an $SO(N)$ symmetric theory with $N$ real scalar fields $\varphi^a$ it reads
\begin{align}\label{eqn:rg improved one loop}
    \frac{\partial}{\partial t} U_k
    &= k \frac{\partial}{\partial k} U_k
    = \frac{Z_k}{2} \int \frac{\dif^d q}{(2 \pi)^d} \, \frac{\partial}{\partial t} P(q)\\
    &\times \biggl(\frac{N - 1}{Z_k P(q) + U_k^\prime} + \frac{1}{Z_k P(q) + U_k^\prime + 2 \rho U_k^{\prime\prime}}\biggr).\notag
\end{align}
Here $U_k$ is a function of $\rho = \frac{1}{2} \varphi^a \varphi_a$ and primes denote derivatives with respect to $\rho$. The integral is ultraviolet and infrared finite due to the properties of the inverse ``average propagator'' $P(q)$,
\begin{equation}\label{eqn:inv avr propagator}
    P(q)
    = \frac{q^2}{1 - f_k^2(q)},
    \quad
    f_k(q)
    = \exp\Bigl[-a (q^2/k^2)^\beta\Bigr].
\end{equation}
It has a simple physical interpretation since for every value of $\rho$ the propagator contains the $\rho$-dependent masses of small fluctuations, i.e. $m^2 = U^\prime$ for the $N - 1$ modes in the ``Goldstone directions'' and $m^2= U^\prime + 2 \rho U^{\prime\prime}$ for the radial mode. If the $k$-dependence of the wave function renormalization $Z_k$ is known (or neglected), \cref{eqn:rg improved one loop} becomes a partial differential equation in the variables $t$ and $\rho$. Its solution for $k \to 0$ provides the effective potential $U(\rho)$ for all values of $\rho$.

Together with a computation \cite{wetterich1993average} of the anomalous dimension $\eta = -\partial_t \ln Z_k$, this equation has produced surprisingly accurate results even for problems without any small coupling. The phase structures of the two- and three-dimensional theories are correctly described \cite{wetterich1993average}, including the Kosterlitz-Thouless phase transition \cite{kosterlitz1973ordering,kosterlitz1974critical} for $d = 2$, $N = 2$. \Cref{eqn:rg improved one loop} also gives an accurate picture \cite{tetradis1993high} of the high temperature phase transition in the four-dimensional theory. The three-dimensional critical exponents relevant for this second-order transition are computed with a surprising precision \cite{tetradis1993high,tetradis1994critical}. Furthermore, the average action describes quantitatively how the effective potential becomes convex for $k \to 0$ in the phase with spontaneous symmetry breaking \cite{ringwald1990average,tetradis1992scale}.

In view of this success the higher loops apparently play no major role. One is led to the question if there might exist an \textit{exact} evolution equation of the type \labelcref{eqn:rg improved one loop}. As a first attempt one may try to use the Schwinger-Dyson equations. These equations involve, however, the bare couplings and it is probably too ambitious to solve an equation which contains the physics at all length scales at once. \Cref{eqn:rg improved one loop} is more modest since only a small momentum range $q^2 \approx k^2$ contributes effectively to the integral. A different possible approach may look for an appropriate truncation of the ``exact renormalization group equation'' derived from an infinitesimal variation of $k$ \cite{wetterich1993average}. This equation is closely related to an equivalent equation which is obtained from the infinitesimal variation of an ultraviolet cutoff in models with explicit momentum cutoff \cite{wegner1976phase,polchinski1984renormalization,ellwanger1994flow}. The difficulty here lies in the fact that the precise properties of $\Gamma_k$ are needed for momenta near the effective cutoff, i.e. $q^2 \approx k^2$. So far we did not find a viable truncation of the infinite system of differential equations involved in this problem.

In this letter we present a new exact evolution equation. It expresses the $k$-dependence of an effective action $\Gamma_k$ in terms of the exact propagator, i.e. the second functional derivative of $\Gamma_k$ with respect to the fields:
\begin{equation}\label{eqn:wetterich}
    \frac{\partial}{\partial t} \Gamma_k[\varphi]
    = \frac{1}{2} \Tr\Bigl[\bigl(\Gamma_k^{(2)}[\varphi] + R_k\bigr)^{-1} \partial_t R_k\Bigr].
\end{equation}
The second functional derivative $\Gamma_k^{(2)}$ and $R_k$ may be expressed either in coordinate space
\begin{alignedeqn}\label{eqn:gamma reg pos}
    &(\Gamma_k^{(2)})_b^a(x,y)
    = \frac{\delta^2 \Gamma_k}{\delta \varphi_a(x) \, \delta \varphi_b(y)},\\
    &(R_k)_b^a(x,y)
    = \delta_b^a \tilde{R}_k(x-y)
    = \delta_b^a \delta(x-y) R_k(i \partial),
\end{alignedeqn}
or in momentum space, where we consider for simplicity a finite volume $\Omega$ (e.g. a torus) with discrete momentum spectrum
\begin{equation}
    \begin{aligned}
        &\varphi^a(x)
        = \sum_q e^{-i q_\mu x^\mu} \varphi^a(q),\\
        &\varphi^a(-q)
        = \varphi_a^\ast(q),
    \end{aligned}
\end{equation}
such that
\begin{equation}\label{eqn:gamma reg mom}
    \begin{aligned}
        (\Gamma_k^{(2)})_b^a(q,q^\prime)
        &= \frac{1}{\Omega }\frac{\partial^2 \Gamma_k}{\partial \varphi_a^\ast(q) \, \partial \varphi^b(q^\prime)},\\
        (R_k)_b^a(q,q^\prime)
        &= \delta_b^a \delta(q-q^\prime) R_k(q).
    \end{aligned}
\end{equation}
Here $R_k(q)$ is the Fourier transform of $\tilde{R}_k(x-y)$, i.e.
\begin{equation*}
    R_k(q)
    = \int \dif^d x \, e^{i q_\mu x^\mu} \tilde{R}_k(x).
\end{equation*}
We assume that it is a positive function of the invariant $q^2 =q^\mu q_\mu$. Its exact shape is arbitrary provided it obeys certain restrictions which will be discussed below. The trace reads in coordinate space
\begin{equation*}
    \Tr = \int \dif^d x \int \dif^d y \sum_a,
\end{equation*}
and in momentum space
\begin{equation*}
    \Tr = \sum_q \sum_a = \Omega \int \frac{\dif^d q}{(2 \pi)^d} \sum_a,
\end{equation*}
where we have taken the limit $\Omega \to \infty$ in the last expression. (The evolution equation \labelcref{eqn:wetterich} can also easily be written for complex scalar fields $\varphi_a(x)$ for which $\varphi^a(q)$ and $\varphi^a(-q)$ are independent variables. In this case $\delta \varphi_a(x)$ should be replaced by $\delta \varphi_a^\ast(x)$ in \cref{eqn:gamma reg pos}. For $m$ complex scalar fields $\Gamma_k^{(2)}$ is considered a $2 m \times 2m$-matrix in internal space, with negative indices defined by $\varphi^{-a}(q) = \varphi_a^\ast(q)$ or $\varphi^{-a}(x) = \varphi_a^\ast(x)$). Our equation \labelcref{eqn:wetterich} has a simple graphical representation which immediately indicates the close analogy to the one-loop equation \labelcref{eqn:rg improved one loop}.
\begin{alignat*}{2}
    &\frac{\partial}{\partial t} \Gamma_k
    &&= \frac{1}{2} \;
    \begin{tikzpicture}[baseline=-3]
        \def\radius{0.5}
        \draw (0,0) circle (\radius);
        \coordinate (cross) at (-\radius,0);
        \draw (cross) node[cross=\radius,thick] {};
        \draw[fill=white,postaction={pattern=north east lines}] (0.9*\radius,0) circle (0.4*\radius);
    \end{tikzpicture}\\
    &\quad\tikz[baseline=-2.5]{\draw node[cross,thick] {};}
    &&= \frac{\partial}{\partial t} R_k\\[1ex]
    &\tikz[baseline=-2.5]{\draw (0,0) -- (1,0);\draw[fill=white,postaction={pattern=north east lines}] (0.5,0) circle (0.2);}
    &&= \text{full $k$-dependent propagator}
\end{alignat*}
The one-loop computation of $\Gamma_k$ becomes exact, but the price to pay is the use of the full propagator! We will see that for a suitable choice of $R_k(q)$ \cref{eqn:rg improved one loop} indeed follows from \labelcref{eqn:wetterich} in a suitable truncation.

In general, the function $R_k(q)$ should obey several requirements, which will be discussed in more detail below. The property
\begin{equation}\label{eqn:infrared reg}
    \lim_{k \to 0} R_k(q)
    = 0
\end{equation}
implies
\begin{equation}\label{eqn:macro limit}
    \lim_{k \to 0} \Gamma_k[\varphi]
    = \Gamma[\varphi],
\end{equation}
where $\Gamma[\varphi]$ is the generating functional for the 1P1 Green functions defined as usual by a Legendre transformation. Secondly, $R_k$ should diverge for $k \to \infty$ and fixed $q^2$. Then one has
\begin{equation}\label{eqn:micro limit}
    \lim_{k \to \infty} \Gamma_k[\varphi]
    = S[\varphi],
\end{equation}
with $S$ the classical action. (In a theory with an effective momentum cutoff $\Lambda$ one has approximately $\Gamma_\Lambda[\varphi] = S[\varphi]$.) The effective action $\Gamma_k$ therefore interpolates between $S$ and $\Gamma$. A solution of the evolution equation \labelcref{eqn:wetterich} for $k \to 0$ with the initial value given by $\Gamma_\Lambda$, permits a computation of $\Gamma$ and in this sense provides a complete solution of the theory. Finally, the momentum integration in \labelcref{eqn:wetterich} should converge rapidly for $q^2 \gg k^2$ and $q^2 \ll k^2$ such that only a small momentum interval $q^2 \approx k^2$ effectively contributes. This can be achieved, for example, by the choice
\begin{equation}
    R_k(q)
    = \frac{\hat{Z}(q) \, q^2 f_k^2(q)}{1 - f_k^2(q)},
\end{equation}
with $f_k$ defined in \cref{eqn:inv avr propagator} and $\hat{Z}(q)$ suitably chosen such that the combination of $R_k$ with the kinetic term $Z_k q^2$ from $\Gamma^{(2)}$ yields approximately the inverse ``average propagator'' \labelcref{eqn:inv avr propagator}
\begin{equation}
    Z_k P(q)
    = Z_k q^2 + R_k.
\end{equation}
With this choice \cref{eqn:wetterich} reduces to \cref{eqn:rg improved one loop} if we approximate $\Gamma_k^{(2)} - Z_k q^2$ by the mass matrix. We emphasize, however, that many alternative choices of $R_k$ remain open. The precise form of the infrared cutoff depends, of course, on the details of the choice for $R_k(q)$.

Let us now prove the exact evolution equation \labelcref{eqn:wetterich} for a theory with $N$ real scalar fields $\chi^a$ and action $S[\chi]$. We add to the kinetic term a piece
\begin{equation}
    \Delta S
    = \frac{1}{2} \Omega \sum_q R_k(q) \chi_a^\ast(q) \chi^a(q),
\end{equation}
and also include sources ($J^a(-q) = J_a^\ast(q)$ for real scalar fields)
\begin{equation}
    S_k
    = S + \Delta S - \sum_q J_a^\ast(q) \, \chi^a(q).
\end{equation}
This defines a $k$-dependent generating functional
\begin{equation}
    W_k[J]
    = \ln Z_k[J]
    = \ln \int \mathcal{D} \chi \, \exp\bigl(-S_k[\chi,J]\bigr).
\end{equation}
In the limit $k \to 0$ $W_k[J]$ becomes the generating functional for the connected Green functions computed with the original classical action $S$ alone. The expectation value of $\chi$ in the presence of $\Delta S$ and $J$ reads
\begin{equation}\label{eqn:ev}
    \varphi^a(q)
    = \langle \chi^a(q)\rangle
    = \frac{\partial W_k}{\partial J_a^\ast(q)},
\end{equation}
and is a functional of the sources $\varphi_k = \varphi_k[J]$. We also define the $k$-dependent connected two-point function
\begin{equation}
    \begin{aligned}
        (G_k)_b^a(q,q^\prime)
        &= \frac{\partial^2 W_k}{\partial J_a^\ast(q) \partial J^b(q^\prime)}\\
        &= \langle \chi^a(q) \chi_b^\ast(q^\prime)\rangle - \langle \chi^a(q)\rangle \langle\chi_b^\ast(q^\prime)\rangle.
    \end{aligned}
\end{equation}
Let us now introduce $\tilde{\Gamma}_k[\varphi_k]$ by a Legendre transformation
\begin{equation}
    \tilde{\Gamma}_k[\varphi_k] + W_k[J] - \sum_q J_a^\ast(q) \varphi_k^a(q)
    = 0,
\end{equation}
such that
\begin{equation}\label{eqn:avr eom}
    \frac{\partial \tilde{\Gamma}_k}{\partial \varphi_k^a(q)}
    = J_a^\ast(q).
\end{equation}
Differentiating \labelcref{eqn:avr eom} with respect to $\varphi_k$ and \labelcref{eqn:ev} with respect to $J^\ast$ yields the well-known identity
\begin{equation}\label{eqn:identity}
    \sum_{q^\prime} (G_k)_b^a(q,q^\prime) \, \frac{\partial^2 \tilde{\Gamma}_k}{\partial \varphi_{kb}^\ast(q^\prime) \, \partial \varphi_k^c(q^{\prime\prime})}
    = \delta_c^a \delta_{q,q^{\prime\prime}}.
\end{equation}
The dependence of $\tilde{\Gamma}_k$ on $k$ for fixed variables $\varphi_k$ is given by
\begin{align}
    \frac{\partial}{\partial t} \tilde{\Gamma}_k\Bigr|_{\varphi_k}\!\!
    &= k \frac{\partial}{\partial k} \tilde{\Gamma}_k\Bigr|_{\varphi_k}
    = -\frac{\partial}{\partial t} W_k\Bigr|_J
    = \frac{\partial}{\partial t} \langle\Delta S\rangle\\
    &= \frac{\Omega}{2} \sum_q \frac{\partial}{\partial t} R_k(q) \bigl[(G_k)_a^a(q,q) + \varphi_{ka}^\ast(q) \varphi_k^a(q)\bigr].\notag
\end{align}
(Note that $J$ is considered here as a $k$-dependent function of $\varphi_k$.) The last term can be absorbed into a redefinition of $\Gamma_k$,
\begin{equation}\label{eqn:definition}
    \Gamma_k
    = \tilde{\Gamma}_k - \frac{\Omega}{2} \sum_q R_k(q) \varphi_{ka}^\ast(q) \varphi_k^a(q).
\end{equation}
This gives the relation
\begin{equation}
    \frac{\partial}{\partial t} \Gamma_k
    = \frac{\Omega}{2} \Tr G_k \frac{\partial}{\partial t} R_k,
\end{equation}
where $G_k$ is regarded as a matrix in internal space and momentum space and $R_k$ is given by \cref{eqn:gamma reg mom}. Inserting \labelcref{eqn:identity} and using the definition \labelcref{eqn:definition} one finally obtains the evolution equation \labelcref{eqn:wetterich}.

In order to understand the role of the scale $k$ in $\Gamma_k$, we first note that the term $\Delta S$ provides an effective infrared cutoff of the order $k$ in the computation of
$W_k$. Similarly the momentum integrations in the computation of $\Gamma_k$ are effectively cut off for $q^2 < k^2$ such that only fluctuations with momenta $q^2 > k^2$ are included in the corresponding functional integral. In short, $\Gamma_k$ is an effective action where the modes with $q^2 > k^2$ have been integrated out. It has all the qualitative properties of the average action \cite{wetterich1991average} for slowly varying fields ($q^2 \ll k^2$). In particular, in the limit $k \to 0$ $\Gamma_k$ tends towards $\tilde{\Gamma}_k$ and $\tilde{\Gamma}_k$ becomes the Legendre transform of $W$. This establishes \cref{eqn:macro limit}. In the opposite limit $k \to \infty$ the quadratic piece $\Delta S$ diverges. The classical approximation to $W_k$,
\begin{equation}
    W_k^{(0)}
    = -S_k\bigl[\chi^0[J]\bigr],
    \quad
    J^\ast
    = \frac{\delta(S + \Delta S)}{\delta \chi}\Bigr|_{\chi^0},
\end{equation}
becomes exact such that $W_k^{(0)}$ is the Legendre transform of $S + \Delta S$. Then $\tilde{\Gamma}_k$ is the Legendre transform of $W_k^{(0)}$ and, since $S + \Delta S$ is always convex in the limit $k \to \infty$,
\begin{equation}
    \Gamma_k
    = \tilde{\Gamma}_k - \Delta S
    = S.
\end{equation}
This proves \cref{eqn:micro limit}. The approximate relation $\Gamma_\Lambda = S$ follows from the good accuracy of the classical approximation for $W_\Lambda$ provided the cutoff $\Lambda$ is sufficiently large compared to all mass scales of the action $S$.

The evolution equation \labelcref{eqn:wetterich} is a partial differential equation for infinitely many variables $t$ and $\varphi^a(q)$. For a practical use one has to find an appropriate truncation. Let us first consider an effective action of the form
\begin{align}\label{eqn:ansatz}
    \Gamma_k
    &= \int \dif^d x \Bigl[U_k(\rho) + \tfrac{1}{2} \partial^\mu \varphi_a \normord{Z_k(\rho,-\partial_\nu \partial^\nu)} \partial_\mu \varphi^a\notag\\
    &\hphantom{{}= \int \dif^d x \Bigl[}+ \tfrac{1}{4} \partial^\mu \rho \normord{Y_k(\rho,-\partial_\nu \partial^\nu)} \partial_\mu \rho\Bigr].
\end{align}
Here the normal ordering $\normord{Z}$ indicates that the derivative operators are always on the right. If $\normord{Z}$ can be expanded around some constant $\rho_0$, it reads
\begin{alignedeqn}
    &\normord{Z_k(\rho,-\partial_\nu \partial^\nu)}\\
    &\hphantom{Z_k}= Z_k(\rho_0,-\partial_\nu \partial^\nu) + (\rho - \rho_0) \, Z_k^\prime(\rho_0,-\partial_\nu \partial^\nu)\\
    &\hphantom{Z_k=}+ \tfrac{1}{2} (\rho - \rho_0)^2 \, Z_k^{\prime\prime}(\rho_0,-\partial_\nu \partial^\nu) + \dots
\end{alignedeqn}
The evolution equation for $U_k$ can now be obtained by expanding around a constant field $\varphi_a(q=0) = \varphi \, \delta_{a1}$, $\rho = \frac{1}{2} \varphi^2$,
\begin{align}
    &\varphi_a(q) = \varphi \, \delta_{a1} + \chi_a(q),\\[2ex]
    &\begin{aligned}
        \Gamma_k
        &= \Omega U_k(\rho) + \Omega \sum_q \bigl(\Gamma_k^{(1)}\bigr)_a(q) \chi^a(q)\\
        &\hphantom{={}}+ \tfrac{1}{2} \Omega \sum_{q,q^\prime} \chi_a^\ast(q) \bigl(\Gamma_k^{(2)}\bigr)_b^a(q,q^\prime) \chi^b(q^\prime) + \dots,
    \end{aligned}
\end{align}
with
\begin{align}
    &\begin{aligned}
        (\Gamma_k^{(2)})_{ab}(q,q^\prime)
        &= M_{ab}^2 \delta_{q,q^\prime} + \Bigl[Z_k(\rho,q^2) \delta_{ab}\\
        &\hphantom{{}=}+ \rho Y_k(\rho,q^2) \delta_{a1} \delta_{b1}\Bigr] q^2 \delta_{q,q^\prime},
    \end{aligned}\\[1ex]
    &\begin{aligned}
        M_{ab}^2
        = \frac{\partial^2 U_k}{\partial \chi^a \, \partial \chi^b}
        = U_k^\prime(\rho) \delta_{ab} + 2 \rho U_k^{\prime\prime}(\rho) \delta_{a1} \delta_{b1}.
    \end{aligned}
\end{align}
We note that the ansatz \labelcref{eqn:ansatz} contains the most general terms for quadratic fluctuations around a constant field (for constant $\rho$ normal ordering is not needed).  The resulting evolution equation for $U_k$ remains therefore exact:
\begin{align}\label{eqn:u flow}
    &\frac{\partial}{\partial t} U(\rho)
    = \frac{1}{2} \int \frac{\dif^d q}{(2 \pi)^d} \, \frac{\partial}{\partial t} R(q) \biggl(\frac{N - 1}{M_0} + \frac{1}{M_1}\biggr),\\[1ex]
    &\begin{aligned}\label{eqn:denominators}
        M_0
        &= Z(\rho,q^2) q^2 + R(q) + U^\prime(\rho),\\
        M_1
        &= \bigl[Z(\rho,q^2) + \rho Y(\rho,q^2)\bigr] q^2 + R(q)\\
        &\hphantom{{}={}}+ U^\prime(\rho) + 2 \rho U^{\prime\prime}(\rho).
    \end{aligned}
\end{align}
Corresponding evolution equations for the derivatives of $U$ are obtained easily by partial differentiation of \cref{eqn:u flow} with respect to $\rho$, i.e.
\begin{alignedeqn}\label{eqn:uprime flow}
    \frac{\partial}{\partial t} U^\prime
    = v_d \int_0^\infty\! \dif x \, x^{\frac{d}{2}-1} \frac{\partial}{\partial t} \Biggl[\frac{(N - 1) (U^{\prime\prime} + Z^\prime x)}{M_0}\\
    + \frac{3 U^{\prime\prime} + 2 \rho U^{\prime\prime\prime} + (Z^\prime + Y + \rho Y^\prime) x}{M_1}\Biggr].
\end{alignedeqn}
Here we have introduced $x = q^2$, $v_d^{-1} = 2^{d+1} \pi^{d/2} \Gamma\bigl(\frac{d}{2}\bigr)$ and it is understood that $\partial/\partial t$ acts only on $R$ on the r.h.s. of \cref{eqn:uprime flow}.

It is advantageous to define
\begin{equation}
    R_k(x)
    = \hat{Z}(x) \, x \, r(x/k^2),
\end{equation}
with
\begin{equation}
    \hat{Z}(x)
    = Z_k(\rho_0(k),x)\bigr|_{k^2=x},
\end{equation}
where $\rho_0(k)$ is the $k$-dependent location of the minimum of $U_k$. \Cref{eqn:rg improved one loop} follows immediately with $Z_k \equiv Z_k(\rho_0(k),k^2)$ and $r = f^2 (1 - f^2)^{-1}$ if we neglect in \labelcref{eqn:u flow} all terms involving $Y_k$ or derivatives of $Z_k$ with respect to $\rho$ or $x$. Since the neglected terms are all related to a scale dependence of the kinetic term, one expects them to vanish for a vanishing anomalous dimension. This justifies the encouraging results of \cref{eqn:rg improved one loop}. Knowing the exact evolution equation \labelcref{eqn:u flow}, one is now able to compute systematically the corrections to \labelcref{eqn:rg improved one loop}. For an approximate solution of \cref{eqn:rg improved one loop} and a computation of $\eta = -\partial_t \ln Z_k$ we refer to refs. \cite{wetterich1993average,tetradis1993high,tetradis1994critical,tetradis1992scale}. We note at this point that for $\beta > 1$ in \cref{eqn:inv avr propagator} the condition \labelcref{eqn:infrared reg} is only fulfilled for $q^2 > 0$ but violated for $q^2 = 0$. The limits $q^2 \to 0$ and $k^2 \to 0$ do not commute. Similarly, for $\beta < 1$ and $\hat{Z}(q^2 \to 0) \to \text{const}$, the function $R_k$ does not diverge for $k \to \infty$ if $q^2= 0$. In order to avoid this complication we emphasize a natural choice in \labelcref{eqn:inv avr propagator}, namely $\beta = 1$, $a = \frac{1}{2}$ such that $P \approx k^2 + \frac{1}{2} q^2$ for $q^2 \ll k^2$. The infrared cutoff contained in P then acts like a mass term, and several of the problems discussed in ref. \cite{wetterich1993average} can be avoided.

In conclusion, we have derived a new exact evolution equation \labelcref{eqn:wetterich} for an effective action $\Gamma_k$. The corresponding system of infinitely many differential equations permits a useful truncation. This allows to compute the effective potential by solving an approximate equation of the type \labelcref{eqn:rg improved one loop}. The momentum integration necessary for the computation of the evolution equation is ultraviolet and infrared finite. The scale $k$ acts like an infrared cutoff in the propagator and the $k$ derivative ensures that large momentum contributions are exponentially suppressed. For momenta $q^2 \ll k^2$ the properties of $\Gamma_k$ are very similar to the average action defined by a constraint in ref. \cite{wetterich1991average}. Since the formulation by a constrained action permits the easy use of field-theoretical techniques beyond the use of evolution equations it would be useful to establish the exact correspondence between the average action and effective action described here.\smallskip

The author thanks M. Reuter and N. Tetradis for collaboration on the average action and many discussions which motivated this work.

\printbibliography

\end{multicols}

\end{document}